\documentclass[aps,pra,showpacs,twocolumn,amsmath,preprintnumbers,nofootinbib,secnumarabic,a4paper,floatfix]{revtex4} 

\usepackage{epsfig,verbatim,hyperref}
\usepackage{amssymb,amsfonts}
\usepackage{graphicx}

\usepackage[caption=false]{subfig}
\def\sct{0.53} 
\def\sct{0.92} 
\def\sct{0.924} 

\def\art{paper}
\def\jrn#1#2#3#4#5#6{\textit{#3} \textbf{#4}, #5 (#6).} \def\boo#1#2#3#4#5#6{\textit{#2} (#3, #4, #5).}    \def\andd{ and } \def\andt{ and } \def\eq{Eq.\,}  \def\eqs{Eqs.\,} \def\Ref{Ref.\,}  

\def\boldsymbol#1{#1}
\def\text#1{{\mbox{#1}}}

\def\eqref#1{(\ref{#1})}

\def\scn#1#2{\section{#1}\lb{#2}}  

\def\bfl{\begin{flushleft}}
\def\efl{\end{flushleft}}
\def\bfr{\begin{flushright}}
\def\efr{\end{flushright}}
\def\bc{\begin{center}}
\def\ec{\end{center}}
\def\be{\begin{equation}}
\def\ee{\end{equation}}
\def\bse{\begin{subequations}}
\def\ese{\end{subequations}}
\def\ba{\begin{eqnarray}}
\def\ea{\end{eqnarray}}
\def\baa#1{\begin{array}{#1}}
\def\eaa{\end{array}}
\def\bw{\begin{widetext}}
\def\ew{\end{widetext}}
\def\nn{\nonumber }
\def\lb#1{\label{#1}}
\def\bit{\begin{itemize}}
\def\eit{\end{itemize}}
\def\bco{}
\def\bcs{\begin{cases}}
\def\ecs{\end{cases}}

\def\lan{{\cal L}}
\def\lanp{{\cal V}}
\def\en{{\epsilon}}

\def\pDer#1#2{\frac{\partial #1}{\partial #2}}

\def\vena{\boldsymbol{\nabla}}

\def\nc0{\tilde b_0}

\def\vol{{\cal V}}
\def\vol{V}

\def\nrmf{{N}}

\def\drm{d}
\def\dvol{\drm\vol}

\def\dn{n}  \def\dnc{\bar{\dn}}

\def\text#1{{\rm #1}}
\def\enk#1{\kappa_{#1}}

\def\res{\nu}
\def\cs0{\bar c_s}

\begin{document}

\preprint{\small Int. J. Mod. Phys. B \textbf{36}, 2250121 (2022) 
\quad [\href{https://doi.org/10.1142/S0217979222501211}{DOI: 10.1142/S0217979222501211}]}


\title{
Resolving the puzzle of sound propagation in a dilute Bose-Einstein condensate 
}

\author{Konstantin G. Zloshchastiev}
\email{https://bit.do/kgz}
\affiliation{Institute of Systems Science, Durban University of Technology, P.O. Box 1334, 
Durban 4000, South Africa\\kostiantynz@dut.ac.za;\\kostya@u.nus.edu}


\begin{abstract} 
A unified model of a dilute Bose-Einstein condensate is proposed, combining of the logarithmic and Gross-Pitaevskii nonlinear terms in a wave equation, where the Gross-Pitaevskii term describes two-body interactions, as suggested by standard perturbation theory;
while the logarithmic term is essentially non-perturbative, and takes into account quantum vacuum effects.
The model is shown to have  excellent agreement with  sound propagation data in the condensate of cold sodium atoms
known since the now classic works by Andrews and collaborators.
The data also allowed us to place constraints on two of the unified model's parameters, which describe the strengths of the logarithmic and Gross-Pitaevskii terms.
Additionally, we suggest an experiment constraining the value of the third parameter (the characteristic density scale of the logarithmic part of the model), using the conjectured attraction-repulsion transition of many-body interaction inside the condensate.
\end{abstract}

\date{received: 8 Dec 2021 [WSPC], 2 Feb 2022 [RG]}


\maketitle

\scn{Introduction}{s:in}
The velocity of the ordinary (``first'') sound in a
magnetically trapped dilute Bose-Einstein condensate of cold sodium atoms at temperatures below a microkelvin 
was measured by Andrews \textit{et al.} \cite{akm97,akm98}
as a function of density 
(where the second paper presented the corrected data).
In essence, the optical
dipole force of a focused laser beam was used to
modify the trapping potential to induce localized excitations. 
The measurements were done in the vicinity of
the center of the Bose-Einstein condensate cloud, where the axial density
varied slowly.
The propagation of sound was observed using rapid sequencing of nondestructive phase-contrast images, and the speed of sound
was plotted as a function of condensate peak density.

\begin{figure}[t]
\centering
\subfloat[data \cite{akm97}]{
  \includegraphics[width=\sct\columnwidth]{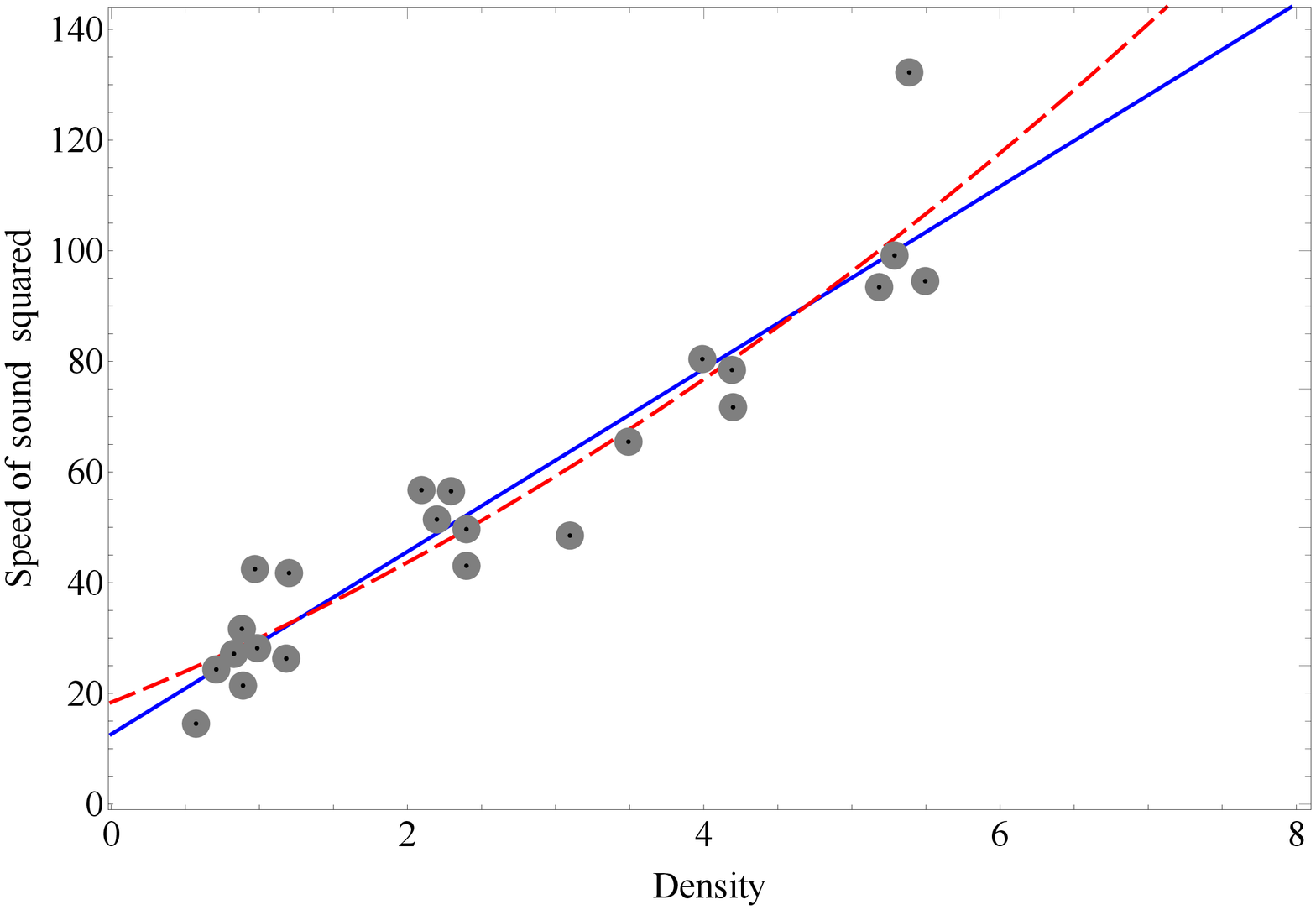}
}
\hspace{0mm}
\subfloat[data \cite{akm98}]{
  \includegraphics[width=\sct\columnwidth]{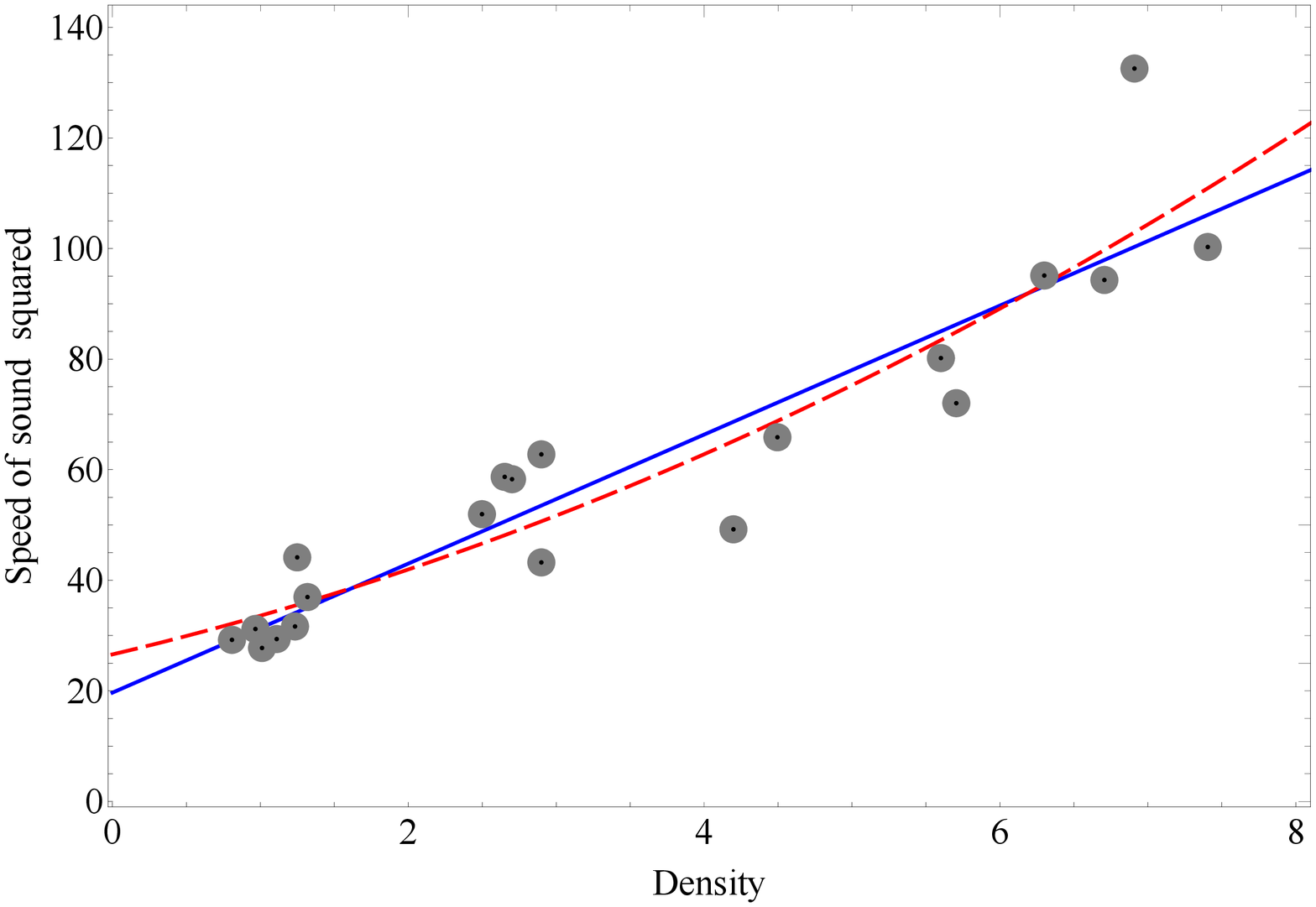}
}
\hspace{0mm}
\subfloat[data \cite{akm97} $\cup$ \cite{akm98}]{
  \includegraphics[width=\sct\columnwidth]{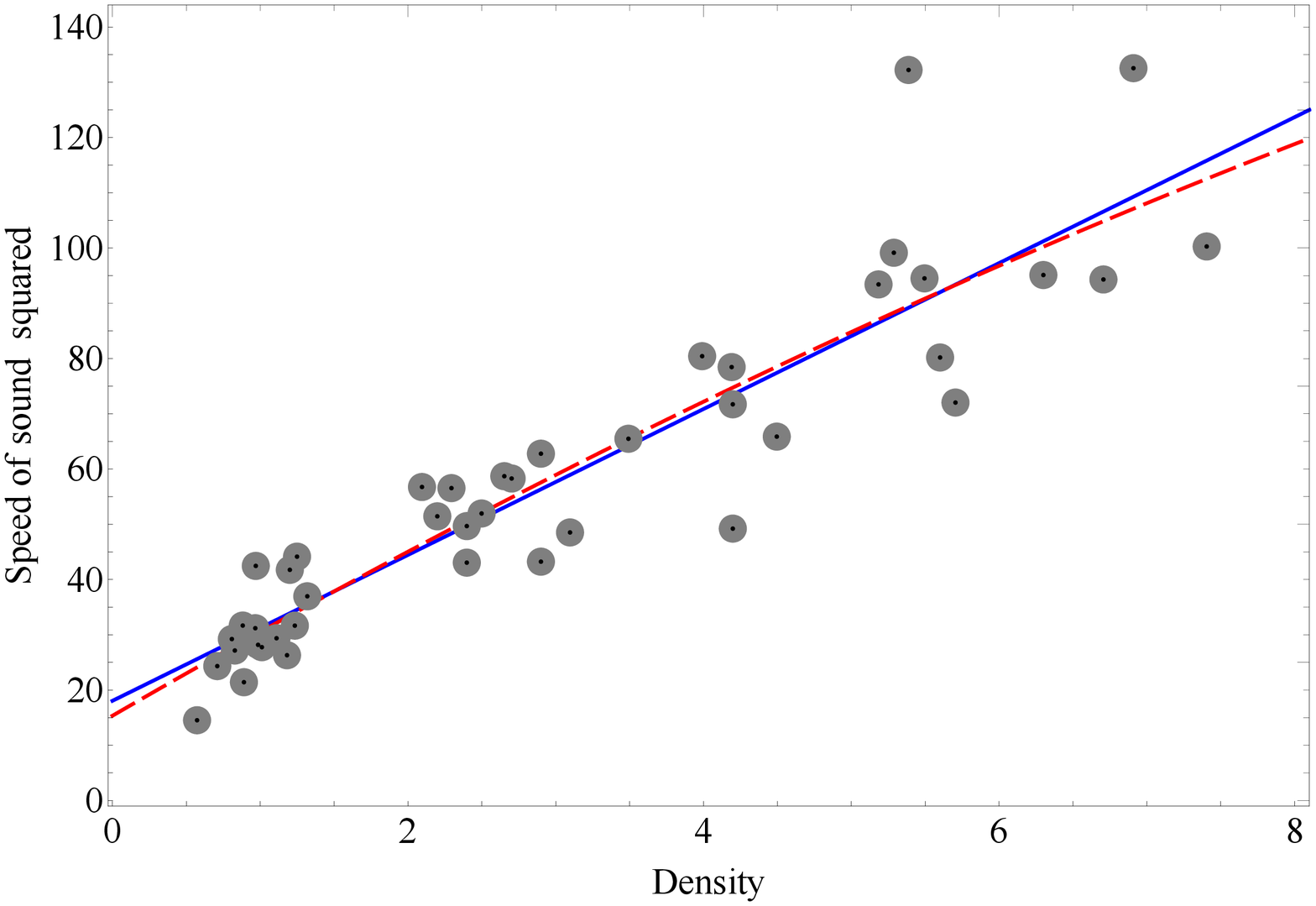}
}
\hspace{0mm}
\caption{Speed of sound squared (in mm$^2\,$s$^{-2}$) versus condensate peak density (in $10^{14}$ cm$^{-3}$), 
for the 
original data set
(\textbf{a}), the
corrected data set
(\textbf{b}), 
and their union set  (\textbf{c}).
The data's mean values are marked by circles, 
the curves are their best fits: linear (solid line) and quadratic (dashed curve).
}
\label{f:fexpsnd}
\end{figure}

A model-unbiased fitting of the 
data 
reveals that sound velocity $c_s$
scales
as a square root of a linear function of density \cite{z21spb}:
\be\lb{e:correct}
c_s  \sim 
\left(n + \res \right)^{1/2}
,
\ee
where $n$ is a particle density of the condensate,
and $\res$ is a positive constant whose value can be phenomenologically estimated  at
between $0.77$ and $4.2\,  \times\! 10^{14}$ cm$^{-3}$, 
depending on the choice of best fit function 
(linear or quadratic) and the data set used (original \cite{akm97} or corrected \cite{akm98}, or a union thereof),
see Fig. \ref{f:fexpsnd} and Table \ref{t:fit}. 
At this stage, we do not assign any physical meaning (such as being related to the particle density)
to the constant $\res$,
but regard it as a phenomenological parameter (which allows multiple interpretations, as we shall see in what follows). 
Similarly, a value of the overall proportionality constant $K$  is estimated 
between $2.5$ and $4.1\, \times\! 10^{-8}$ cm$^{5/2}$ s$^{-1}$.

Note that although two types of best fit functions (linear and quadratic) were considered for velocity of sound squared in the work \cite{z21spb},
the quadratic function is used only for confirmation of a non-zero value of $\res$ being a systematic effect but not an artifact of the fitting procedure.
Other than that, one can exclusively deal with linear best fit functions in dilute Bose-Einstein condensates,
because the higher-order terms in density come from the multiple-body (three of more) scattering's perturbative corrections
which are usually neglected for low-density condensates at low temperatures.

In the formula above,
the occurrence of ``residual'' sound velocity 
$ 
\cs0 \equiv \lim\limits_{n\to\,0} c_s = \sqrt\res K
$, 
is surprising;
because conventional theory of dilute Bose-Einstein condensates,
based on the Gross-Pitaevskii (GP) equation alone \cite{gr61,pi61,gor21,psbook}, predicts 
a simpler behaviour $c_s  \sim n^{1/2}$
(in other words, $\cs0$ would be identically zero in the GP theory).
At first look, the difference between these two behaviours is insignificant.
However, a nonzero $\cs0$ is not just a curious puzzle; it also raises a profound interpretation problem:
if the speed of sound does not vanish when the condensate density 
decreases down to an infinitesimal value,
then what is the physical nature of the matter which remains as $n \to 0$, 
as compared to a case when no condensate exists ($n \equiv 0$).

An intuitive qualitative answer to this puzzle can be imagined in terms of virtual particles
and zero-point fluctuations of the condensate,
which is somewhat analogous to the well-known quantum-mechanical phenomenon of 
non-vanishing energy of a quantum harmonic oscillator in the limit when the number of its modes tends to zero.
However,
the quantitative description of this phenomenon is more complex, because it requires a model which
goes beyond the perturbation theory in general or Gross-Pitaevskii approximation in particular.
A working example of such a model in a theory of quantum Bose liquids can be found in \Ref \cite{sz19},
but the case of a dilute Bose gas requires some adaption of that framework.

In this \art, we propose a 
non-perturbative model, which takes into account not only two-body interactions, but also
vacuum effects in Bose-Einstein condensates of alkali atoms,
and analytically explains the empirical formula \eqref{e:correct}.
In section \ref{s:mod} we describe a model of a dilute Bose-Einstein condensate,
which is expected to resolve the above-mentioned puzzle.
In section \ref{s:snd} we derive the model's equation of state and hence speed of sound, reproduce the
formula \eqref{e:correct}, and set experimental bounds for the model's parameters.
Conclusions are drawn in section \ref{s:con}.

\scn{The model}{s:mod}
We begin by introducing a complex-valued condensate wavefunction
$\Psi = \Psi (\textbf{x},t)$,
which 
obeys a normalization condition
$ 
\int_\vol |\Psi|^2 \dvol  
= \int_\vol \dn \, \dvol 
= \nrmf 
$, 
where 
$\vol$ and $\nrmf$
are, respectively, the total volume and number of particles  of the condensate,
$\dn = |\Psi|^2$,  
and $m$ 
is the mass of the constituent particle (in this case, an alkali atom) \cite{psbook}.

Aside from this condition, the condensate wavefunction
obeys a wave equation, which is defined based on a chosen model.
We select the model, which is a special case
of the one proposed in \Ref \cite{sz19} to describe superfluid helium.
Some simplification is possible in the case of the system \cite{akm97,akm98},
because their cold atom clouds have a much lower density than superfluid helium -- 
which 
allows us to neglect
multiple-body (three or more) interactions.
Therefore,
in the equations (3) -- (6) of \Ref \cite{sz19}, 
one can assume the maximal interaction multiplicity number to be ${\cal N} \equiv 2$.
Under these assumptions, 
we obtain the following  wave equation:
\be\lb{e:becgeneq}
\left[
- i \hbar \, \partial_t
- \frac{\hbar^2}{2 m} \vec \nabla^2
+
V_{\text{ext}} (\vec x,\, t)
+
F(|\Psi|^2)
\right]
\Psi
= 0,
\ee
where 
$V_{\text{ext}}$ is an external/trapping potential, 
and function $F$ is defined as
a sum of two terms
\be
F(\dn) \equiv
F_{(\text{ln})} (\dn) 
+
F_{(2)} (\dn) 
=
\enk{(\text{ln})}
\ln{\!\left(
\dn/\dnc
\right)}
+
\enk{(2)}
\dn
,
\lb{e:F} 
\ee
where
we denoted
$
\enk{(\text{ln})} \equiv \en C_{(\text{ln})}
$,
$
\enk{(2)} \equiv -2 \en C_{(2)}/\dnc
$,
and
$\en$ and $\dnc$ are scale constants with the dimensions of energy and particle density, respectively,
whereas 
$C$'s are dimensionless coupling coefficients
(in notations used in \Ref \cite{sz19}).
For the purposes of this paper, in the leading-order approximation 
the external potential can be neglected, 
hence we assume the trapless condensate $V_{\text{ext}} \equiv 0$ in what follows,
which is a robust approximation for systems of such type \cite{kp97,z21ijmpb}.


\bw\bc
\squeezetable
\begin{table}[t] 
\resizebox{\textwidth}{!}{
\begin{tabular}{|c|ccccc|ccccc|}
\hline\hline
\multicolumn{1}{|c|}{~\textbf{Source~~}}&\multicolumn{5}{c|}{\textbf{Linear best fit}}&\multicolumn{5}{c|}{\textbf{Quadratic best fit}}\\
\cline{2-11} 
\textbf{of}&
~ $\nu$ &~ $K$ &~ $\enk{(\text{ln})}/m$ &~ $\enk{(2)}/m$&~$\cs0$&
~ $\nu$ &~ $K$ &~ $\enk{(\text{ln})}/m$ &~ $\enk{(2)}/m$&~$\cs0$  \\
\textbf{data}&
~$(10^{14} \frac{1}{\text{cm}^3})$&~($10^{-8} \frac{\text{cm}^{5/2}}{\text s}$)&~($\frac{\text{cm}^{2}}{\text s^{2}}$)&~($10^{-15} \frac{\text{cm}^{5}}{\text s^{2}}$)&~($\frac{\text{cm}}{\text s}$)&
~$(10^{14} \frac{1}{\text{cm}^3})$&~($10^{-8} \frac{\text{cm}^{5/2}}{\text s}$)&~($\frac{\text{cm}^{2}}{\text s^{2}}$)&~($10^{-15} \frac{\text{cm}^{5}}{\text s^{2}}$)&~($\frac{\text{cm}}{\text s}$)\\
[1mm]
\hline
\cite{akm97}&
0.77&4.06 &0.13 & 1.65& 0.36&
1.72&3.27 &0.18 & 1.07& 0.43\\
\cite{akm98}&
1.69&3.42 &0.20 & 1.17& 0.44&
4.22&2.51 &0.27 & 0.63& 0.52\\
\cite{akm97} $\cup$ \cite{akm98}&
1.37&3.63 &0.18 & 1.32& 0.42&
0.99&3.94 &0.15 & 1.55& 0.39\\
[1mm]
\hline\hline
\end{tabular}
} 
\caption{Experimental bounds for parameters of the logarithmic-quartic model \eqref{e:becgeneq}, \eqref{e:F},
based on the simplest polynomial best fit functions for $c_s^2$ data, cf. Fig. \ref{f:fexpsnd}.
The lines refer to, from top to bottom, the original data set \cite{akm97},
the corrected data set \cite{akm98}, and a union thereof.
The quadratic best fit is used only for confirmation of a non-zero value of $\res$ being a systematic effect,
therefore it can be disregarded for the purposes of this \art. 
}\label{t:fit}
\end{table}
\ec\ew

According to these equations,
our Bose-Einstein condensate 
is expected to have a two-part structure (we use the term ``part'' here to differentiate 
from physically two-component fluids or mixtures models
which use separate wave functions for each component):

First, the part described by the 
term $F_{(2)} \left(|\Psi|^2\right)
\propto |\Psi|^2 $,
is responsible for the Gross-Pitaevskii 
two-body contact interaction discussed above.

The other term 
$F_{(\text{ln})} \left(|\Psi|^2\right)  \propto
\ln{\left(
|\Psi|^2/\dnc
\right)}
$
describes the logarithmic fluid part in the wave equation.
Nonlinearity of this type often occurs in theories containing quantum Bose liquids and condensates
\cite{z11appb,az11,z12eb,z17zna,z19ijmpb,sz19,z20un1,rc21,z21ltp,z21ijmpb}.
The reason for such universality is that  logarithmically nonlinear terms readily emerge in evolution equations for those dynamical systems in which interparticle interaction energies dominate kinetic ones, see \Ref \cite{z18zna} for more details.

Since the works by Rosen and Bialynicki-Birula and Mycielski \cite{ros68,bb76}, 
mathematical properties of (purely) logarithmic nonlinear wave equations have been extensively studied, 
to mention only very recent literature 
\cite{zsj21,itw21,ji21,liu21,liu21fp,ll21,pww21,hl21,pj21,bcs21,cf21,la21,psw21,kj22,sav22,yzq22,aea21,zsw22,pjh22,vgm22,cs22nu,pen22};
but combined logarithmic-polynomial nonlinearities, 
like the one occurring in our model, still await thorough study.

\scn{Equation of state and speed of sound}{s:snd}
Using the Madelung representation of a wavefunction,
one can rewrite  any nonlinear wave equation of type \eqref{e:becgeneq} in hydrodynamic form.
One can 
derive the general expressions for
an equation of state and speed of sound $c_s$,
which become quite simple if we retain only the leading-order terms with respect to the Planck constant
\cite{z19mat}.
When evaluated on a function \eqref{e:F}, 
those formulae 
yield,
respectively,
an equation of state and speed of sound for our model:
\ba
P
&\approx& 
\int \dn F'(\dn) \drm \dn
=
\enk{(\text{ln})} \dn
+
\enk{(2)}
\dn^{2}/2
, \lb{e:eos0}\\
c_s^2 
&\equiv&
\frac{1}{m} \pDer{P}{n}
\approx
\frac{1}{m}
\left(
\enk{(\text{ln})} 
+
\enk{(2)}
\dn
\right)
,
\lb{e:cs0}
\ea
where the approximation symbol
means that we kept only the leading-order terms with respect to
the Planck constant.
In the derived equation of state, one can notice the Bogoliubov term, which 
is quadratic in density and induced by the Gross-Pitaevskii part of our model, cf. \eq \eqref{e:F}.
This is an expected effect of the GP model \cite{psbook}.
However, 
one can also notice the linear term, which is well-known from ideal gas models.
It is interesting
that here it comes from the logarithmic part of our model.

Furthermore, formula \eqref{e:cs0} can be written as
\ba
c_s 
&\approx&
\sqrt{
\frac{
\enk{(2)}}{m}
}
\left(
\dn
+
\frac{\enk{(\text{ln})} }{\enk{(2)}}
\right)^{1/2}
,
\lb{e:cs0exp}
\ea
which reproduces the empirical formula \eqref{e:correct},
once we make the following associations:
\be\lb{e:assoc}
\nu \Leftrightarrow 
\enk{(\text{ln})} /\enk{(2)},
\ \
K^2 \Leftrightarrow
\enk{(2)}/m 
,
\ee 
thus one can see that it is the logarithmic nonlinearity 
which induces 
the constant term $\nu$ in an expression for the speed of sound squared.
Inverting these formulae, we obtain the following relations between parameters of our model and 
experimentally measured values:
\be\lb{e:exp}
\enk{(\text{ln})} = m\, \nu K^2, \ \
\enk{(2)} = m K^2
,
\ee
which can be used to place empirical bounds on the model's parameters.

\begin{figure}[t]
\begin{center}
\includegraphics[width=1.02\columnwidth]{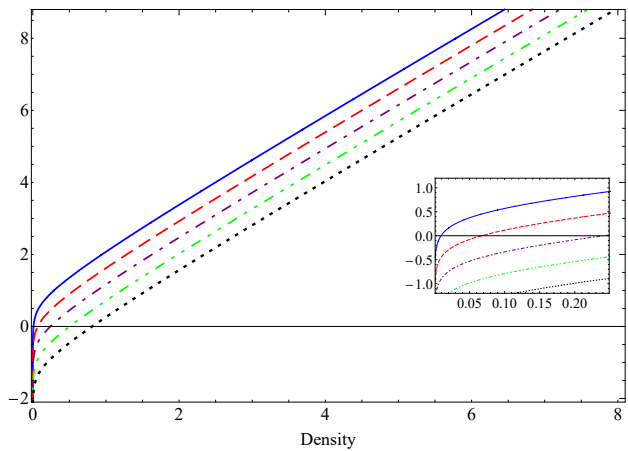}
\end{center}
\caption{Profile of $F (\dn)/m$ (in cm$^2\,$s$^{-2}$), 
versus condensate density $n$ (in $10^{14}$ cm$^{-3}$), 
for different scales of $\dnc$ (in $10^{14}$ cm$^{-3}$): 0.01 (solid), 0.1 (dashed), 1 (dash-dotted), 10 (dash-double-dotted) and 100 (dotted).  
Values of $\kappa$'s are taken from the linear best bit of data \cite{akm98}, see Table \ref{t:fit}. 
}
\label{f:fpot}
\end{figure}

In Table \ref{t:fit}, 
we show these constraints, based on linear and quadratic best fit functions of $c_s^2$ data, as described in Fig. \ref{f:fexpsnd}.
These are placed on the parameters $\enk{(\text{ln})}$ and $\enk{(2)}$ of the model \eqref{e:becgeneq}, \eqref{e:F},
but not on the parameter $\dnc$ which determines the characteristic density scale of logarithmic nonlinearity.
This parameter is thus assumed to be a free parameter of our model for now,
but one can think of various ways of how to determine its value 
in future experiments.

For example,
one can think of an experiment which is sensitive to the transition of function $F (n)$
between its 
positive and negative values'
regimes
(the existence of such a transition is theoretically predicted, as it can be seen directly from Fig. \ref{f:fpot}). 
If such a transition occurs, the critical density value $n_c$ exists at which $F (n)$ momentarily turns to
zero before flipping its sign.
Then the density scale parameter of the logarithmic nonlinearity would be related to this value 
as
\be
\dnc = n_c
\exp{
\left(
\enk{(2)} n_c/\enk{(\text{ln})} 
\right)
}
=
n_c 
\exp{
\left(
n_c/\res
\right)
}
,
\ee
according to \eqs \eqref{e:F} and \eqref{e:exp}.

\scn{Conclusion}{s:con}
To summarize, we formulated a two-part Bose-Einstein condensate 
model, involving both logarithmic and Gross-Pitaevskii nonlinearities,
which is a truncated version of the model previously used by us for superfluid helium.
In this model, the Gross-Pitaevskii two-body interaction term is suggested by the
perturbation theory of dilute Bose-Einstein condensates;
whereas the logarithmic term is essentially non-perturbative, and takes into account quantum vacuum effects.
Our model has excellent agreement with 
the sound propagation data in the condensate of cold sodium atoms by Andrews \textit{et al.},
where the logarithmic term induces a shift constant in the asymptotic behaviour of sound velocity at
infinitesimal values of density, while the
Gross-Pitaevskii term defines a shape of the curve itself.
The data also allowed us to place constraints for two of the model's parameters, which describe 
the strength of the logarithmic and Gross-Pitaevskii terms.
Furthermore,
we suggested an experiment constraining the value of the third parameter (the density scale of the logarithmic component of the model); based on the conjectured attraction–repulsion transition  
of the resulting many-body interaction in the condensate.\\~\\~\\~\\


\begin{acknowledgments}
This work is based on the research supported by the 
Department of Higher Education and Training of South Africa (University Research Outputs Support programme),
and
in part by the 
National Research Foundation of South Africa (Grants Nos. 95965, 131604 and 132202). 
Proofreading of the manuscript by P. Stannard is greatly appreciated.
\end{acknowledgments}



\def\PR{Physical Review}
\def\JMP{Journal of Mathematical Physics}
\def\CMP{Communications in Mathematical Physics}
\def\APNY{Annals of Physics}
\def\PL{Physics Letters}
\def\PSc{Physica Scripta}
\def\IJTP{Int. J. Theor. Phys.}
\def\GC{Gravitation \& Cosmology}
\def\APP{Acta Physica Polonica}
\def\ZN{Zeitschrift f\"ur Naturforschung}
\def\JPB{Journal of Physics B: Atomic, Molecular and Optical Physics}
\def\EPJ{European Physical Journal}
\def\EPL{Europhysics Letters (EPL)}
\def\GAFD{Geophysical \& Astrophysical Fluid Dynamics}
\def\ARMA{Archive for Rational Mechanics and Analysis}
\def\ARFM{Annual Review of Fluid Mechanics}
\def\JPCS{Journal of Physics: Conference Series}

\end{document}